\documentclass[aps,twocolumn,floatfix,superscriptaddress,preprintnumbers,showpacs,showkeys]{revtex4}
\usepackage{amssymb}
\usepackage{amsmath}
\usepackage{amsfonts}
\usepackage{epsfig}
\usepackage{graphicx}
\usepackage{tabularx}

\newcommand{\bea}{\begin{eqnarray}}
\newcommand{\ena}{\end{eqnarray}}
\newcommand{\nn}{\nonumber\\}
\newcommand{\be}{\begin{equation}}
\newcommand{\en}{\end{equation}}

\newcommand{\ed}{\end{document}}

\newcommand{\slp}{p\kern-5pt/}

\begin{document}

MITP/15-085 (Mainz) 

\title{Towards an assessment of the ATLAS data on \\
the branching ratio $\Gamma(\Lambda_b^0 \rightarrow 
\psi(2S)\Lambda^0)/\Gamma(\Lambda_b^0 \rightarrow J/\psi\Lambda^0)$} 

\author{Thomas Gutsche}
\affiliation{Institut f\"ur Theoretische Physik, Universit\"at T\"ubingen,
Kepler Center for Astro and Particle Physics, 
Auf der Morgenstelle 14, D-72076, T\"ubingen, Germany}

\author{Mikhail A. Ivanov}
\affiliation{Bogoliubov Laboratory of Theoretical Physics, 
Joint Institute for Nuclear Research, 141980 Dubna, Russia}

\author{J\"{u}rgen G. K\"{o}rner}
\affiliation{PRISMA Cluster of Excellence, Institut f\"{u}r Physik, 
Johannes Gutenberg-Universit\"{a}t, 
D-55099 Mainz, Germany}

\author{Valery~E.~Lyubovitskij}
\affiliation{Institut f\"ur Theoretische Physik, Universit\"at T\"ubingen,
Kepler Center for Astro and Particle Physics, 
Auf der Morgenstelle 14, D-72076, T\"ubingen, Germany}
\affiliation{ 
Department of Physics, Tomsk State University,  
634050 Tomsk, Russia} 
\affiliation{Mathematical Physics Department, 
Tomsk Polytechnic University, 
Lenin Avenue 30, 634050 Tomsk, Russia} 

\author{Pietro Santorelli}
\affiliation{
Dipartimento di Fisica, Universit\`a di Napoli
Federico II, Complesso Universitario di Monte Sant' Angelo,
Via Cintia, Edificio 6, 80126 Napoli, Italy} 
\affiliation{
Istituto Nazionale di Fisica Nucleare, Sezione di Napoli, 
80126 Napoli, Italy}

\today

\begin{abstract}

Recently the ATLAS Collaboration at CERN reported on the measurement of 
the branching ratio $R_{\Lambda_b} = \Gamma(\Lambda_b^0 \rightarrow 
\psi(2S)\Lambda^0)/\Gamma(\Lambda_b^0 \rightarrow J/\psi\Lambda^0)$. 
The measured branching ratio $R_{\Lambda_b} = 0.501 \pm 0.033$ (stat) 
$\pm 0.019$ (syst) was found to be lower than the covariant quark 
model prediction of $R_{\Lambda_b} = 0.8 \pm 0.1$  
calculated by us recently. 
We present a detailed analysis of the branching ratio $R_{\Lambda_b}$ 
using a model-independent framework for the heavy-to-light form factors 
based on results from our previos papers. 

\end{abstract}


\maketitle

\section{Introduction}

Recently the ATLAS Collaboration at CERN reported on the measurement of
the branching ratio $R_{\Lambda_b} = \Gamma(\Lambda_b^0 \rightarrow           
\psi(2S)\Lambda^0)/\Gamma(\Lambda_b^0 \rightarrow 
J/\psi\Lambda^0)$~\cite{Aad:2015msa}.
The measured branching ratio $R_{\Lambda_b} 
= 0.501 \pm 0.033$ (stat) $\pm 0.019$ (syst)
was found to be lower than the expectation from the covariant quark 
model calculation~\cite{Ivanov:1996pz,Ivanov:1996fj,Ivanov:1999bk,%
Gutsche:2013pp,Gutsche:2013oea,Gutsche:2014zna,Gutsche:2015mxa} 
$R_{\Lambda_b} = 0.8 \pm 0.1$ done by us recently~\cite{Gutsche:2013oea}. 
Note, in Ref.~\cite{Gutsche:2013oea} 
we have only listed the central value of
$R_{\Lambda_b} = 0.8 \pm 0.1$. The error on the branching ratio was 
determined using several overall fits to a wide spectrum of data on heavy 
hadron decays all with similar good $\chi^{2}$ values. The set of fit values
lead to branching ratios within 0.1 deviation from 0.8.  
We mention that there have been a number of theoretical
quark model calculations for the decay $\Lambda_b \to \Lambda + J/\psi$
based on the factorization
hypothesis~\cite{Cheng:1995fe}-\cite{Mott:2011cx} two of which we will return
to when we present our numerical results. 

In Ref.~\cite{Gutsche:2013oea} we have presented a detailed analysis 
of the branching ratio $R_{\Lambda_b}$ using covariant quark model 
for the heavy-to-light form factors based on results 
from our previos papers. 
In particular, we have shown that our model transition form factors 
(see Ref.~\cite{Gutsche:2014zna} for heavy-to-light transitions and 
Ref.~\cite{Gutsche:2015mxa} for heavy-to-heavy transitions) 
can be approximated to a high accuracy by the double-pole representation 
\bea\label{doublepole_formula}  
F(q^2) = \frac{F(0)}{1- a q^2/M_1^2 + b q^4/M_1^4} \,, 
\ena 
where $M_1$ is the mass of the initial baryon and $a$ and $b$ are fit 
parameters. In the examples studied by us  
we noticed that the fit parameters are approximately related by
$b \approx a^{2}/4$, i.e. the $q^2$ behavior of our 
form factors is very close to a dipole form 
\bea\label{dipole_formula} 
F(q^2) = \frac{F(0)}{(1-q^2/M_d^2)^2} 
\ena 
where $M_d$ is the dipole mass. The scale set by this mass is close 
to the value of the $B_s$ meson mass $M_{B_s} = 5.367$~GeV in the case 
of the $b \to s$ transitions. 

The main objective of the present paper is to show that the explicit 
value of the dipole mass is crucial to understand the branching ratio 
$R_{\Lambda_b}$. Our main result is that $M_d \sim M_{B_s}$ 
leads to $R_{\Lambda_b} \sim 1$, 
while increasing/decreasing values of $M_d$ lead to decreasing/increasing 
values of $R_{\Lambda_b}$. 
In particular, the central ATLAS value $R_{\Lambda_b}  \sim 0.5$ 
would require a relatively large value of the dipole mass $M_d \sim 10$~GeV. 

We start with the definition of the transition amplitude of the process 
$B_1(p_1)\to B_2(p_2) + W_{\rm off-shell}(q)$ which are described by 
the vector and axial vector current matrix elements 
$M_\mu^{V/A}(\lambda_{1},\lambda_{2}) = \langle B_2,\lambda_{2}|J_\mu^{V/A}|
B_1,\lambda_{1}\rangle$. The matrix elements 
can be expanded in terms of a complete set of invariants: 
\bea
M_\mu^V(\lambda_{1},\lambda_{2}) 
&=& \bar u_2(p_2,\lambda_{2})\bigg[F_1^V(q^2)\gamma_\mu
  -\frac{F_2^V(q^2)}{M_1}
  i\sigma_{\mu q}\nonumber\\
&+&\frac{F_3^V(q^2)}{M_1}q_\mu\bigg]u_1(p_1,\lambda_{1}) 
\label{eq: invariants}
\ena
and  
\bea
M_\mu^A(\lambda_{1},\lambda_{2}) 
&=& \bar u_2(p_2,\lambda_{2})
\bigg[F_1^A(q^2)\gamma_\mu-\frac{F_2^A(q^2)}{M_1}
  i\sigma_{\mu q}\nonumber\\
  &+&\frac{F_3^A(q^2)}{M_1}q_\mu\bigg]\gamma_{5}
u_1(p_1,\lambda_{1}) 
\ena
where $M_1$ and $M_2$ are the masses of the initial and final baryon, 
$\sigma_{\mu q} = 
\frac{i}{2}(\gamma_\mu\!\!\not\!\! q\, -\! \not\!\! q\,\gamma_\mu)$ 
and  $q = p_1 - p_2$. The labels $\lambda_{i}=\pm \frac12$ denote the 
helicities of the two baryons. For completeness we have also included the 
form factors $F_{3}^{V/A}$ even though they do not contribute to the
process $\Lambda_{b} \to \Lambda\,+\,J/\psi,\psi(2S)$. They would determine 
the rate for the decay $\Lambda_{b} \to \Lambda\,+\,\eta_{c}$. 

In the heavy quark limit (HQL) the matrix element 
for the heavy-to-light $b \to s$ transition is given 
in terms of two ($f_1$, $f_2$) form factors 
\cite{Hussain:1990uu,Mannel:1990vg,Hussain:1992rb}. The HQL form factors 
depend 
on the variable $p_2\cdot v_{1}$, where $v_{1}=p_{1}/M_{1}$ is the 
four-velocity of the $\Lambda_b$\,. The matrix element now reads  
\bea 
M_\mu^{V-A}= \bar u(p_2) \Big( f_1(p_2v) 
+ \not\! v f_2(p_2v) \Big) O_\mu u(v)  
\ena 
where $O_\mu = \gamma_\mu (1-\gamma^5)$. In the HQL we have used 
the heavy ($b$ quark) mass expansion for the $\Lambda_b$ 
mass~\cite{Gutsche:2013oea}
\bea 
M_{\Lambda_b} = m_b + \bar\Lambda + {\cal O}(1/m_b) 
\ena 
and keep the two leading terms in the expansion --- the heavy quark mass 
$m_b$ and the so-called 
binding energy $\bar\Lambda = {\cal O}(m_b^0)$. The value 
$\bar\Lambda = 0.53$ GeV is fixed using experimental values for 
$M_{\Lambda_b} = 5.6194$ GeV~\cite{Agashe:2014kda} 
and a model value for the constituent mass of the $b$ quark $m_b = 5.09$ GeV.
The constituent mass of the $b$ quark was fixed from an analysis of a wide 
spectrum of data on heavy hadron decays in our approach. 

In the HQL the six form factors $F_{1,2,3}^{V/A, {\rm HQL}}$ 
become related to the HQL form factors $f_{1,2}$ 
as follows~\cite{Hussain:1990uu,Mannel:1990vg,Hussain:1992rb}
\bea 
F_1^{V, {\rm HQL}}& =& F_1^{A, {\rm HQL}} 
 = f_1 + \frac{M_2}{M_1} \, f_2\,, \\
F_2^{V, {\rm HQL}}  &=& F_2^{A, {\rm HQL}} = 
 - F_3^{V, {\rm HQL}} = - F_3^{A, {\rm HQL}}=-f_{2}\nonumber   
\ena 
It is convenient to analyze the decay $\Lambda_b \to \Lambda + V$ 
in terms of the helicity amplitudes $H^{V/A}_{\lambda_2\lambda_V}$ which are 
linearly 
related to the invariant form factors $F_i^{V/A}$ (see details 
in Refs.~\cite{Gutsche:2013pp,Gutsche:2013oea,Gutsche:2014zna,Gutsche:2015mxa})
\be
H^{V/A}_{\lambda_2\lambda_V} = 
M_\mu^{V/A}(\lambda_2)\epsilon^{\dagger\,\mu}(\lambda_V)\,, 
\label{eq:hel_def}\,‚
\en
where $\lambda_V$ is the helicity of the vector meson.  
From angular momentum conservation, one has 
$\lambda_{1}=-\lambda_{2}+\lambda_{W}$. 

The helicity amplitudes read 
(see e.g. Refs.~\cite{Gutsche:2013pp,Gutsche:2013oea,Gutsche:2014zna,Gutsche:2015mxa})
\bea
\nn
H_{+\frac12 +1}^{V/A}&=&\sqrt{2Q_\mp}
  \bigg(F_1^{V/A}\pm \frac{M_\pm}{M_1}F_2^{V/A}\bigg),
\nn
H_{+\frac12 0}^{V/A}&=&\frac{\sqrt{Q_\mp}}{\sqrt{q^2}}
  \bigg(M_\pm F_1^{V/A}\pm \frac{q^2}{M_1} F_2^{V/A}\bigg)\,, 
\label{eq:hel_inv}
\ena
where we make use of the abbreviations $M_\pm = M_1\pm M_2$ and 
$Q_\pm = M_\pm^2  - q^2$.
From parity or from an explicit calculation, one has 
$H_{-\lambda_2,-\lambda_V}^{V/A}  = \pm H_{\lambda_2,\lambda_V}^{V/A}$. 
The total left--chiral helicity amplitude is defined by the composition
\be 
H_{\lambda_2,\lambda_V} = H_{\lambda_2,\lambda_V}^V 
- H_{\lambda_2,\lambda_V}^A \,.
\en 
The weak nonleptonic decays $\Lambda_b \to \Lambda + J/\psi$ and 
$\Lambda_b \to \Lambda + \psi(2S)$ 
are described by bilinear forms of the helicity amplitudes termed
helicity structure functions. The relevant bilinear forms for the rate are 
${\cal H}_U   = |H_{+\frac12 +1}|^2 + |H_{-\frac12 -1}|^2$   
(unpolarized transverse) and 
${\cal H}_L    = |H_{+\frac12\, 0}|^2 + |H_{-\frac12\, 0}|^2$   
(longitudinal) where the rate is proportional to 
${\cal H}_U+{\cal H}_L:={\cal H}_{U+L}$.   
In the HQL the two helicity structure functions can be expressed 
in terms of the functions $f_1$ and $f_2$ as 
\bea 
{\cal H}_U &=& 4 \Big[ (Q_+ + Q_-) \, (f_1^2 + f_2^2) 
+ 8 M_1 M_2 f_1 f_2 \Big]
\,,\nonumber\\
{\cal H}_L &=& {\cal H}_S \ = \ \frac{2}{q^2} \, 
\Big[ Q_+ \Big( M_- \Big[f_1 + \frac{M_2}{M_1} f_2\Big] 
+ \frac{q^2}{M_1} f_2\Big)^2 
\nonumber\\
&+&
Q_- \Big( M_+ \Big[f_1 + \frac{M_2}{M_1} f_2\Big] 
- \frac{q^2}{M_1} f_2\Big)^2 \Big]
\ena 
The $\Lambda_b \to \Lambda + V$ decay rate is given by 
\bea 
\Gamma(\Lambda_b \to \Lambda + V) &=& 
\frac{G_F^2}{32 \pi}  \, \frac{|{\bf p}_V|}{M_1^2} \,   
|V_{cb} V_{cs}^\ast|^2 \, 
C_{\rm eff}^2 \, f_V^2 \, M_V^2  \nonumber\\ 
&\times& {\cal H}_{U+L} \,. 
\ena 
where $|{\bf p}_V| = \sqrt{Q_+ Q_-}/(2 M_1)$ 
is the three-momentum of the decay products in the rest frame of the parent 
baryon, 
$C_{\rm eff} = - 0.262$ is a combination of the relevant 
Wilson coefficients, $f_V$ is the decay constant ($f_{J/\psi} = 415$ MeV, 
$f_{\psi(2S)} = 295.6$ MeV) of the respective vector meson. 

The branching ratio $R_{\Lambda_b}$ can be written in terms of a model
independent factor $R_M$ given by 
\bea 
R_M = 
\frac{|{\bf p}_{\psi(2S)}|}{|{\bf p}_{J/\psi}|} \, 
\biggl(\frac{M_{\psi(2S)} f_{\psi(2S)}}{M_{J/\psi} f_{J/\psi}} \biggr)^2 = 
0.535025 
\ena 
and a model dependent factor $R_{\cal H}$ given by the ratio of the  
helicity structure functions ${\cal H}_{U+L}$ evaluated at 
$q^{2}=M^{2}_{\psi(2S)}$ and $q^{2}=M^{2}_{J/\psi}$
\bea 
R_{\cal H} = \big({\cal H}_{U +L}\big)_{\psi(2S)}\,/\,
\big({\cal H}_{U +L}\big)_{J/\psi}
\ena  
such that
\begin{equation}
R_{\Lambda_b}=R_{M}\,\cdot\,R_{\cal H} 
\end{equation}
Note that the branching ratio $R_{\Lambda_b}$ does not depend on the flavor
and color dependent product of coefficients
$V_{cb} V_{cs} \, C_{\rm eff}$.

Next we need to analyze the $q^{2}$--dependence of the ratio $R_{\cal H}$. 
According to the ATLAS data $R_{\cal H}$ must be close to 1 which implies a 
weak dependence of the helicity structure function on $q^2$ in the range 
between $q^2 = M_{J/\psi}^2 \simeq 9.591$~GeV$^2$ and 
$q^2 = M_{\psi(2S)}^2 \simeq 13.587$~GeV$^2$. 
Our recent analysis showed that due the dipole-like 
behavior of the form factors (\ref{dipole_formula}) with $M_d \simeq M_{B_s}$ 
there is a sizeable growth of the rate structure function 
${\cal H}_{U +L}$ from $q^2 = M_{J/\psi}^2$ to $q^2 = M_{\psi(2S)}^2$. 
Quantitatively, the rapid growth of ${\cal H}_{U +L}$ between 
$q^2 = M_{J/\psi}^2$ to $q^2 = M_{\psi(2S)}^2$ follows from the details of our
dynamical quark model ansatz. Qualitatively, the growth of ${\cal H}_{U +L}$
results from the simple picture of a dipole behavior of the form factors
characterized by the mass scale $\Lambda \simeq M_{B_s}$. The scale corresponds
to the flavor composition of a $t$--channel meson exchange. 
Below we show details of our numerical analysis. In particular, we expose 
the dependence of $R_{\cal H}$ on the dipole mass $M_d$ and show that 
the ATLAS result can be reproduced only with a relatively large value of 
the dipole mass $M_d \sim 10$ GeV which exceeds the mass scale set by the
$B_s$ meson mass by $86\,\%$. 

We want to emphasize that the dipole approximation is also a very 
good approximation for the $q^{2}$--dependence of the $b \to c$ transition 
form factors in the decay
$\Lambda_{b}\to\Lambda_{c}\,\ell^{-}\,\bar\nu_{\ell}$~\cite{Gutsche:2015mxa}. 
Again the dipole mass of 
$M_d = 6.46-6.58$~GeV  is close 
to the ($b \bar c$) mass scale of 6.28 GeV set by the $B_c$ meson mass. 

The details of the calculations of the covariant form factors for the  
$\Lambda_b \to \Lambda$ transition can be found in 
Ref.~\cite{Gutsche:2013oea}. In particular, we used the following 
set of the constituent quark masses  
$m_b = 5.09$ GeV, $m_s = 0.424$~GeV, $m_u = m_d = 0.235$ GeV 
and the actual value of $f_{\psi(2S)} = 286.7$~MeV. 
 
In Figs.~1-6 we present plots of the $q^2$--dependence of our form factors 
from $q^2=0$ to $q^2_{\rm max} = (M_{\Lambda_b}-M_{\Lambda})^2$. 
The two sets of Figs.~1-3 and 4-6 corresponds to the form factors $F_1^{V}$ 
and $F_2^{V}$, respectively. 
In particular, in Figs.~1 and 4 we compare the form factors 
in the exact case and in HQL. The leading $F_1^{V}$ form 
factors in the exact and in the HQL case can be seen to be very close 
to each other. In Figs.~2 and 5 we present a 
comparison of the 
exact form factors (dotted line), and the double-pole  
(curve 1) and dipole approximations 
(curve 2) to the form factors. An analogous comparison in the HQL is presented 
in Figs.~3 and 6. 
One can see that the double-pole approximation lies practically on top of the
exact form factors. Also the dipole approximation can be seen to be quite
a reasonable approximation to both. The parameters for the double-pole and 
dipole approximations are 
summarized in Tables~1-4. We do not list the corresponding values for the
form factor $F_2^A$ since $F_2^A$ is suppressed. 

In Table 5 we present our results for the $\Lambda_b \to \Lambda + V$ 
branching ratios and the ratio $R_{\Lambda_b}$ considering different limiting
cases: 1) exact, 2) exact taking 
into account only the leading form factors $F_1^V$ and $F_1^A$, 3) HQL and 
4) HQL taking 
into account only the leading form factors 
$F_1^{V, {\rm HQL}} = F_1^{A, {\rm HQL}}$. 
One can see that the restriction to the leading form factors $F_1^V$ and 
$F_1^A$ gives a qualitatively reasonable approximation for the evaluation 
of the branching ratios $B(\Lambda_b \to \Lambda + V)$.   

In Fig.~7 we plot the dependence of the ratios 
$R_{\cal H}$ and $R_{\Lambda_b}$ on the dipole mass using an
approximation where $F_{2}^{V/A}=0$
and where the dominating form factors $F_1^V$ and $F_1^A$ have a dipole
form. The plot
encapsules the essence of the results found in Ref.~\cite{Gutsche:2013oea}: 
for a dipole mass of $M_d \sim M_{B_s}$, which reflects the typical scale 
for the $b \to s$ baryon transitions, one finds $R_{\cal H} \sim 2$. Such a 
large value of $R_{\cal H}$ leads to a branching ratio 
$R_{\Lambda_b} \sim 1$ which exceeds the measured central 
value of $R_{\Lambda_b} \sim 0.5$~\cite{Aad:2015msa}. The ATLAS result can
only be reproduced with $M_d \sim 10$ GeV which, as remarked on before, 
is far away from the mass scale set by the
$B_s$ meson mass. 

It is interesting to compare our results with the predictions of
Refs.~\cite{Wei:2009np,Mott:2011cx} who have conveniently supplied 
parametrizations of their model form factors.
The predictions of both models~\cite{Wei:2009np,Mott:2011cx} for the
branching fraction
$B(\Lambda_b \to \Lambda + J/\psi)$ are close to our result. However,
the predictions for the branching fraction 
$B(\Lambda_b \to \Lambda + \psi(2S))$ considerably differ from our result
leading to very different results on the branching ratio $R_{\Lambda_b}$.
The authors of~\cite{Wei:2009np} obtain $R_{\Lambda_b}=0.65$, while the
authors of~\cite{Mott:2011cx} obtain a value of $R_{\Lambda_b}$ exceeding 1.  
The result of Ref.~\cite{Wei:2009np} means that the value of their effective 
dipole mass is much larger than predicted by our approach. 
                                                                       
For completeness we also present predictions of our model for 
analogous mesonic decays 
$B \to K + J/\psi, \psi(2S)$ using the 
covariant quark model results given in~\cite{Ivanov:2011aa}. 
The $B \to K + J/\psi, \psi(2S)$ decay widths are calculated according to 
the formula 
\bea 
\hspace*{-.9cm}
\Gamma(B \to K + V) &=&
\frac{G_F^2}{16 \pi}  \, \frac{|{\bf p}_V|}{M_1^2} \,
|V_{cb} V_{cs}^\ast|^2 \,
C_{\rm eff}^2 \, f_V^2 \, M_V^2 \, |H_{0}|^2 \nonumber\\
&=& 
\frac{G_F^2}{4 \pi}  \, |{\bf p}_V|^3 \,
|V_{cb} V_{cs}^\ast|^2 \,
C_{\rm eff}^2 \, f_V^2 \, f_+^2(M_V^2)\,, 
\ena
where $H_{0}$ is the scalar helicity amplitude given by~\cite{Ivanov:2011aa} 
\bea\label{H0_BK} 
H_0 = \frac{2M_1 |{\bf p}_V|}{M_V} \, f_+(M_V^2) \,.
\ena 
The form factor $f_+(M_V^2)$ multiplies the Lorentz-structure $(p_1+p_2)^\mu$
and has to be evaluated for $q^{2}=M_V^2$\,.  
The $B\,\to\, K$ transition form factors calculated in~\cite{Ivanov:2011aa} 
are very close to a monopole formula
\bea\label{monopole_formula}
f_+(q^2) = \frac{f_+(0)}{1-q^2/M_m^2}
\ena
where $M_m$ is the monopole mass. In the case of the
mesonic $b \to s$ transitions the monopole mass $M_{m}$ is of the order 
of 5 GeV. As before the branching ratio  
$R_B = \Gamma(B \rightarrow \psi(2S) K)/\Gamma(B \rightarrow J/\psi K)$ 
can be written in terms of the two factors 
$R_B = R_M R_{\cal H}$, where $R_M = 0.515792$ and $R_{\cal H}$ depends 
of the transtion form factor $f_+$ according to Eq.~(\ref{H0_BK}). 
In Fig.~8 we display the dependence of the ratios 
$R_{\cal H}$ and $R_{B}$ on the value of the monopole mass $M_m$.
Due to the monopole form of the mesonic transition form factors the 
dependence of $R_{\cal H}$ and thereby $R_{B}$ on the monopole
mass is weaker than in the baryon case. 
In particular, 
$R_B$ changes from 0.39 to 0.24 when $M_m$ changes from 5 to 10 GeV. 
The exact value $R_{B}$ predicted by our approach is $R_{B} = 0.34$. 
This value is smaller than the experimental data on $R_B$:  
0.611 $\pm$ 0.019 (fit)
and 0.603 $\pm$ 0.021 (average)~\cite{Agashe:2014kda}. The smaller ratio 
predicted 
by our model is due to the smaller branching for 
the $B \rightarrow \psi(2S) K$ mode ${\rm Br}(B \rightarrow \psi(2S) K) =  
3.5 \times 10^{-4}$, 
while the experimental value is 
$(6.27 \pm 0.24) \times 10^{-4}$ (fit) and 
$(6.5 \pm 0.4) \times 10^{-4}$ (average)~\cite{Agashe:2014kda}.   
On the other hand, one should mention that our prediction for 
${\rm Br}(B \rightarrow \psi(2S) K)$ is close   
to the results of the {\it BaBar} Collaboration 
$(4.9 \pm 1.6 \pm 0.4) \times 10^{-4}$~\cite{Aubert:2005vi}.  
In case of the $B \rightarrow J/\psi K$ mode we calculate 
${\rm Br}(B \rightarrow J/\psi K) = 9.6 \times 10^{-4}$, which is    
in good agreement with data  
$(10.27 \pm 0.31) \times 10^{-4}$ (fit) and 
$(10.24 \pm 0.35) \times 10^{-4}$ (average)~\cite{Agashe:2014kda}.    

The cases of 
nonleptonic meson and baryon $b \to s$ transitions appear to differ only by  
the power-scaling of the transition form factors. 
The different powers in the form factors (monopole in the meson sector and 
dipole in the baryon sector) leads to smaller values of the branching
ratio in the meson sector, i.e. one has $R_B < R_{\Lambda_b}$. 
We hope that future experiments will clear up the picture on the branching 
ratios $R_{\Lambda_b}$ and $R_B$. 

\begin{table}
\caption{Parameters for the double-pole approximation 
for our form factors $F_1^V$, $F_2^V$ and $F_1^A$.} 
\begin{center}
\def\arraystretch{1}
\begin{tabular}{cccc}
\hline
       &\qquad $F_1^V$ \qquad & \qquad $F_2^V$ \qquad &
        \qquad $F_1^A$ \qquad \\
\hline
$f(0)$ & 0.107  & 0.043   & 0.104   \\
$a$    & 2.271  & 2.411   & 2.232   \\
$b$    & 1.367  & 1.531   & 1.328   \\
\hline
\end{tabular}
\label{tab:1}
\end{center}

\caption{Parameters for the dipole approximation for our form factors 
$F_1^V$, $F_2^V$ and $F_1^A$.} 
\begin{center}
\def\arraystretch{1}
\begin{tabular}{ccccc}
\hline
       &\qquad $F_1^V$ \qquad & \qquad $F_2^V$ \qquad &
        \qquad $F_1^A$ \qquad \\ 
\hline
$f(0)$          & 0.107   & 0.043   & 0.104  \\
$M_d$ (GeV)     & 5.445   & 5.286   & 5.484  \\
\hline
\end{tabular}
\label{tab:2}
\end{center}

\caption{Parameters for the double-pole approximation 
for our form factors 
$F_1^{V, {\rm HQL}}$ and $F_2^{V, {\rm HQL}}$ in HQL.} 
\begin{center}
\def\arraystretch{1}
\begin{tabular}{ccc}
\hline
       &\qquad $F_1^{V, {\rm HQL}}$ \qquad 
       &\qquad $F_2^{V, {\rm HQL}}$  \qquad \\
\hline
$F(0)$ & 0.124  & 0.023   \\
$a$    & 2.259  & 2.464   \\
$b$    & 1.360  & 1.597   \\
\hline
\end{tabular}
\label{tab:3}
\end{center}

\caption{Parameters for the dipole approximation for our form factors 
$F_1^{V, {\rm HQL}}$ and $F_2^{V, {\rm HQL}}$ in HQL.} 
\begin{center}
\def\arraystretch{1}
\begin{tabular}{ccc}
\hline 
&\qquad $F_1^{V, {\rm HQL}}$ \qquad & \qquad $F_2^{V, {\rm HQL}}$  \qquad \\
\hline
$F(0)$          & 0.107   & 0.043  \\
$M_d$ (GeV)     & 5.458   & 5.240  \\
\hline
\end{tabular}
\label{tab:4}
\end{center}

\begin{center}
\caption{Branching ratios 
$B(\Lambda_b \to \Lambda + V)$ \\
(in units of $10^{-4}$)  
and ratio $R_{\Lambda_b}$}

\vspace*{.1cm}

\def\arraystretch{1}
    \begin{tabular}{|c|c|c|c|c|}
      \hline
Quantity                              & Exact & Exact & HQL   & HQL \\
                                      &       & $F_{2}^{V/A}=0$  
                                      &       & $F_{2}^{V/A, {\rm HQL}}=0$ \\
\hline 
$B(\Lambda_b \to \Lambda + J/\psi)$   & 8.90 & 7.34 & 9.55  & 10.27 \\ 
$B(\Lambda_b \to \Lambda + \psi(2S))$ & 7.25 & 5.96 & 7.40  & 8.52  \\
$R_{\Lambda_b}$                       & 0.81 & 0.81 & 0.77  & 0.83  \\
\hline
\end{tabular}
\label{tab:5}
\end{center}

\begin{center}
\caption{Comparison of our results for branching ratios
$B(\Lambda_b \to \Lambda + V)$ (in units of $10^{-4}$)
and ratio $R_{\Lambda_b}$ with approaches~\cite{Wei:2009np,Mott:2011cx}
using their parametrization for form factors.}

\vspace*{.1cm}

\def\arraystretch{1}
    \begin{tabular}{|c|c|c|c|c|}
      \hline
Quantity
& Our & Ref.~\cite{Wei:2009np} & Ref.~\cite{Mott:2011cx}  \\
\hline
$B(\Lambda_b \to \Lambda + J/\psi)$ & 8.90  & 8.44 & 8.21 \\
$B(\Lambda_b \to \Lambda + \psi)$   & 7.25  & 5.48 & 9.35 \\
$R_{\Lambda_b}$                     & 0.81  & 0.65 & 1.14 \\
\hline
\end{tabular}
\label{tab:6}
\end{center}
\end{table}

Let us summarize the main results of our paper. 
Using the covariant quark model we have analyzed 
the branching ratio $R_{\Lambda_b} = \Gamma(\Lambda_b^0 \rightarrow 
\psi(2S)\Lambda^0)/\Gamma(\Lambda_b^0 \rightarrow J/\psi\Lambda^0)$ 
which was recently measured by the ATLAS Collaboration  
CERN~\cite{Aad:2015msa}: 
$R_{\Lambda_b} 
= 0.501 \pm 0.033$ (stat) $\pm 0.019$ (syst). The measurement disagrees with
our 
prediction $R_{\Lambda_b} = 0.8 \pm 0.1$ calculated by us in 
Ref.~\cite{Gutsche:2013oea} by 
2.8 standard deviations. 
The error bars include variation of the 
quark masses $m_b = 5.068 \pm 0.022$~GeV, $m_s = 0.426 \pm 0.002$~GeV and 
$m_u = m_d = 0.238 \pm 0.003$~GeV and decay constant 
$f_{\psi(2S)}$ from 286.7 MeV (old value) to 295.6 MeV (updated value). 
The same variation of the parameters and $f_{\psi(2S)}$ in the meson case 
gives $R_B = 0.35 \pm 0.05$. 
In the present paper we have presented arguments supporting our  
considerations on 
the dipole-like 
behavior of the leading $\Lambda_b \to \Lambda$ transition form factors 
characterized by a dipole mass $M_d$ close to the mass of the $B_s$ meson. 
Such dipole-like behavior is universal not only for heavy-to-light, but 
also for heavy-to-heavy transitions. In the two cases the values of the 
dipole masses are 
found to be very close to the masses of the $(q_{1}\bar q_{2})$ mesons
that are active in the $q_{1} \to q_{2}$ current induced transition. 
In particular, 
the values of the dipole mass
$M_d$ found for the $b \to c$ and $b \to s$ 
transitions are close to the $B_c$ and $B_s$ meson masses, respectively. 
It holds in both limiting cases --- for finite values of the 
heavy quark masses and in the HQL. 

\begin{acknowledgments}

This work was supported by the Tomsk State University Competitiveness 
Improvement Program and the
Russian Federation program ``Nauka'' (Contract No. 0.1526.2015, 3854). 
M.A.I. acknowledges the support of the Mainz Institute for Theoretical 
Physics (MITP). M.A.I. and J.G.K. thank the Heisenberg-Landau Grant for
support.  

\end{acknowledgments}

\clearpage 

\begin{figure}
\begin{center}
\epsfig{figure=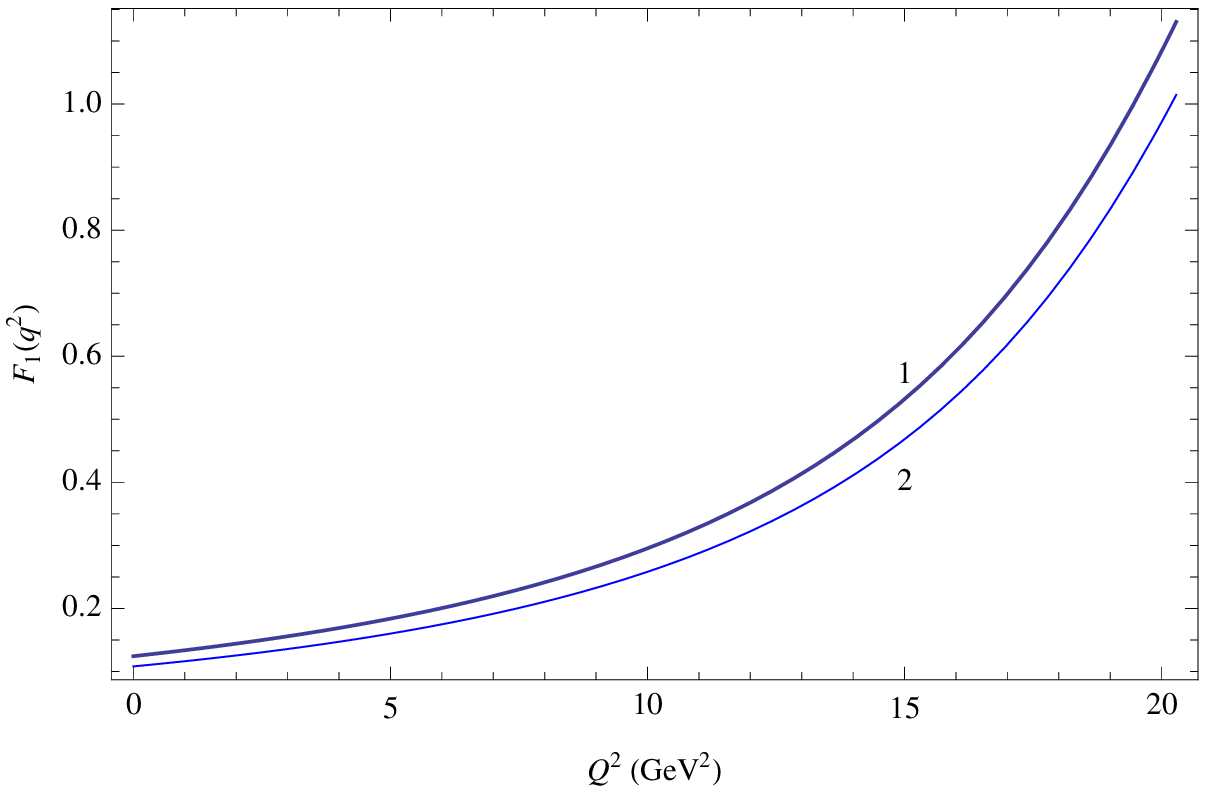,scale=.65}
\end{center}
\vspace*{-.6cm}
\noindent
\caption{Form factors $F_1^{V, {\rm HQL}}$ (curve 1) 
and $F_1$ (curve 2).
\label{fig1}}   

\begin{center}
\epsfig{figure=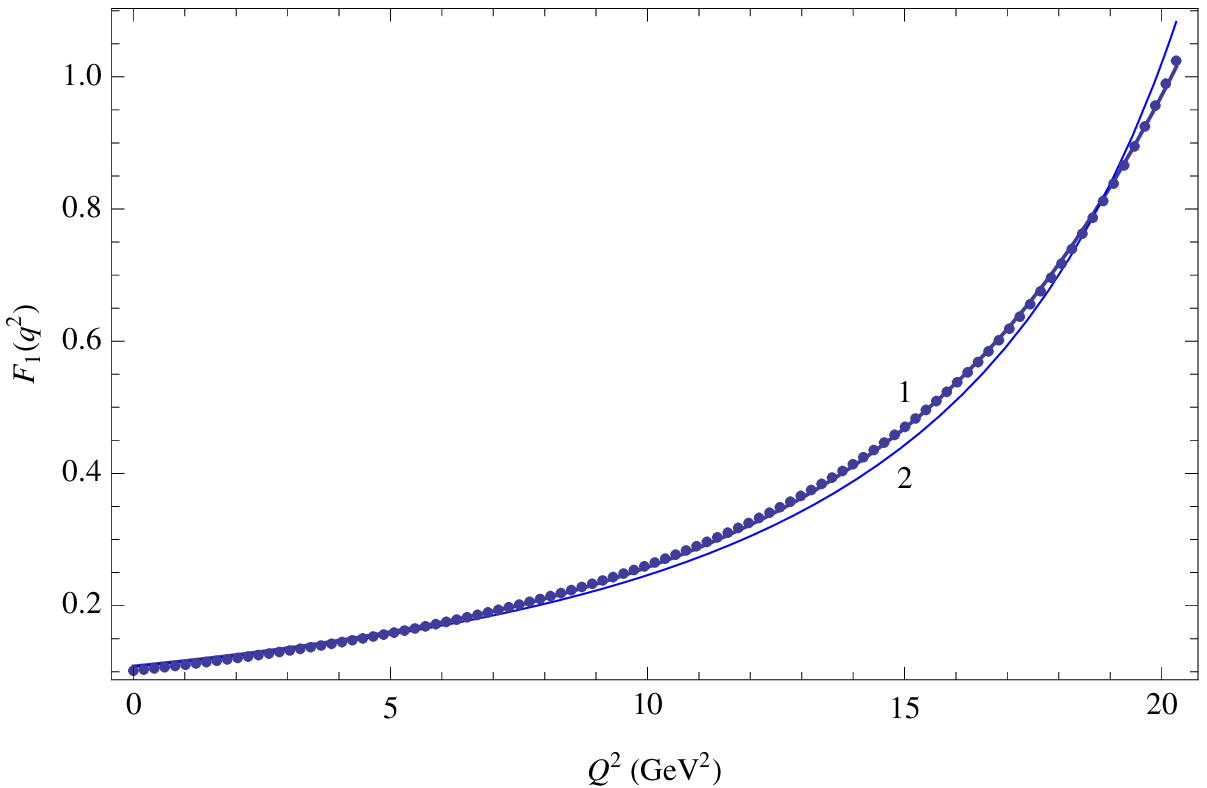,scale=.65}
\end{center}
\vspace*{-.6cm}
\noindent
\caption{Form factor $F_1^{V}$: exact result (dotted line), \\{}
double-pole approximation (curve 1) and 
dipole approximation (curve 2).
\label{fig2}}   

\begin{center}
\epsfig{figure=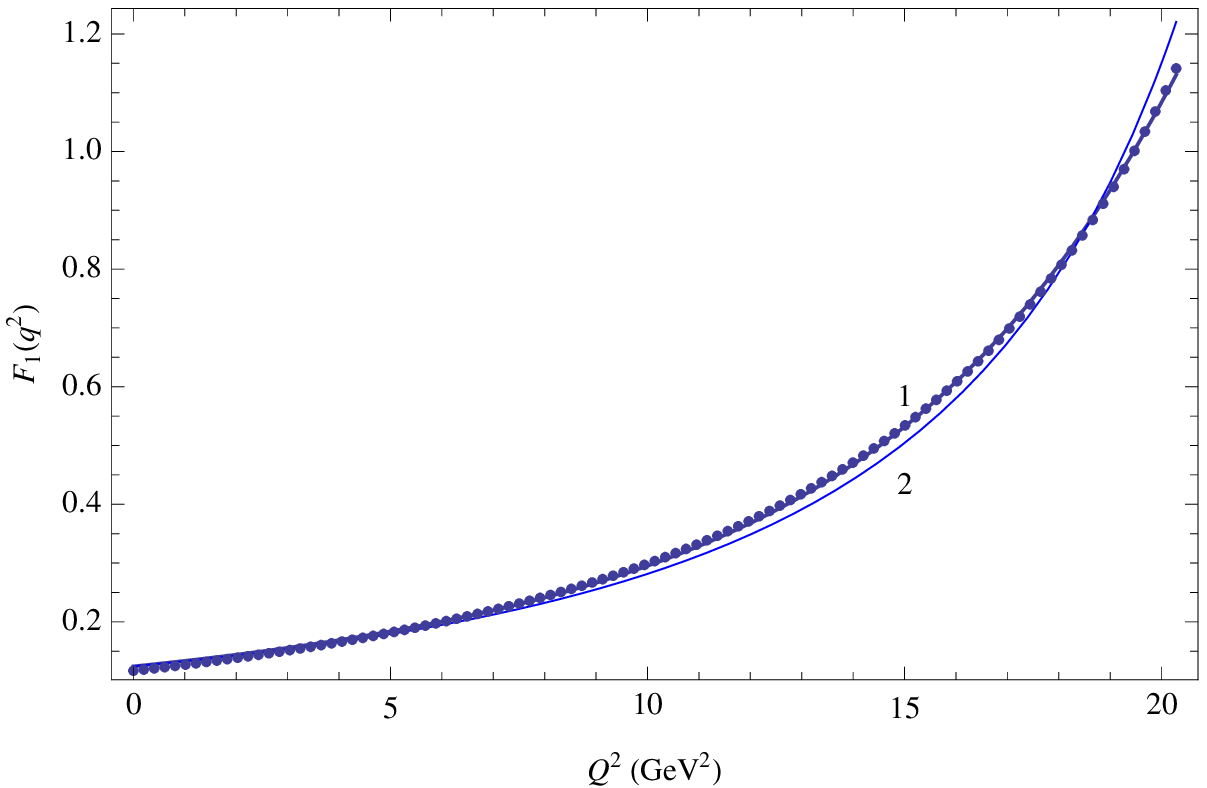,scale=.65}
\end{center}
\vspace*{-.6cm}
\noindent
\caption{Form factor $F_1^{V, {\rm HQL}}$: exact result (dotted line), \\{} 
double-pole approximation (curve 1) and 
dipole approximation (curve 2).
\label{fig3}}   
\end{figure}

\begin{figure}
\begin{center}
\epsfig{figure=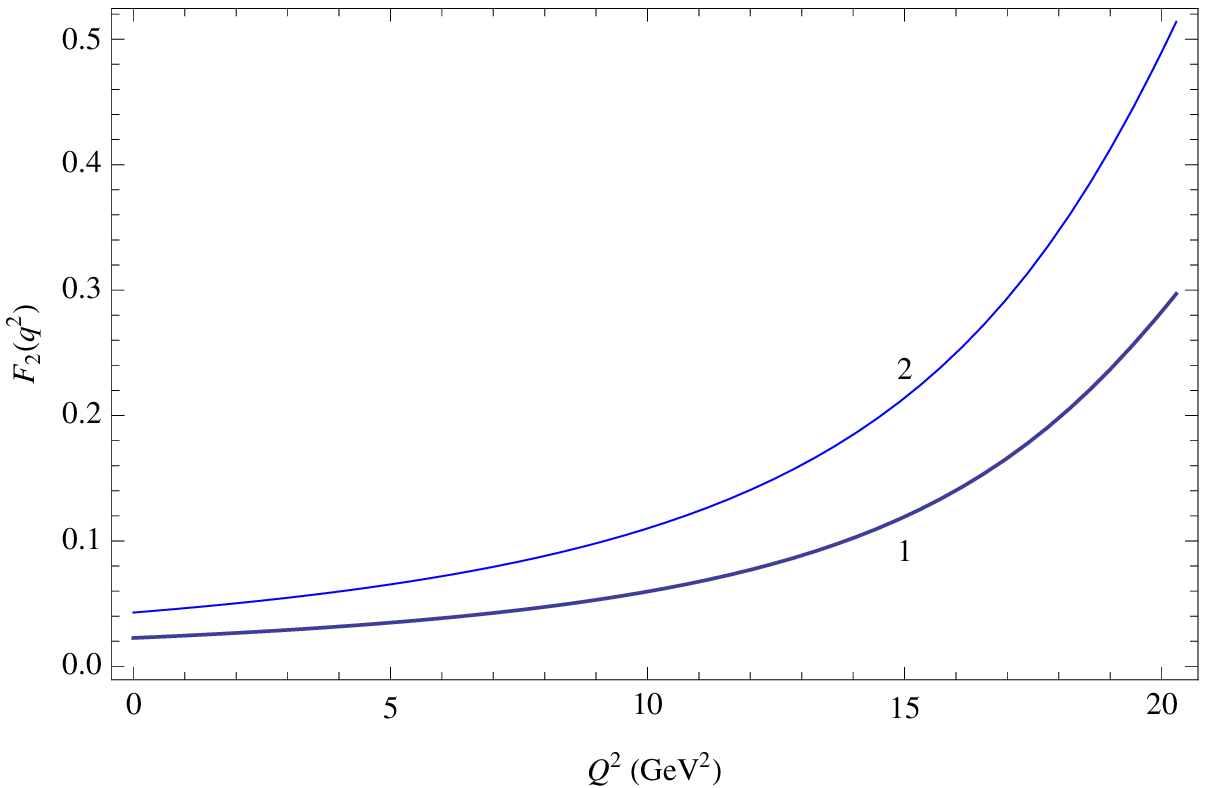,scale=.65}
\end{center}
\vspace*{-.6cm}
\noindent
\caption{Form factors $F_2^{V, {\rm HQL}}$ (curve 1) 
and $F_2$ (curve 2).
\label{fig4}}   

\begin{center}
\epsfig{figure=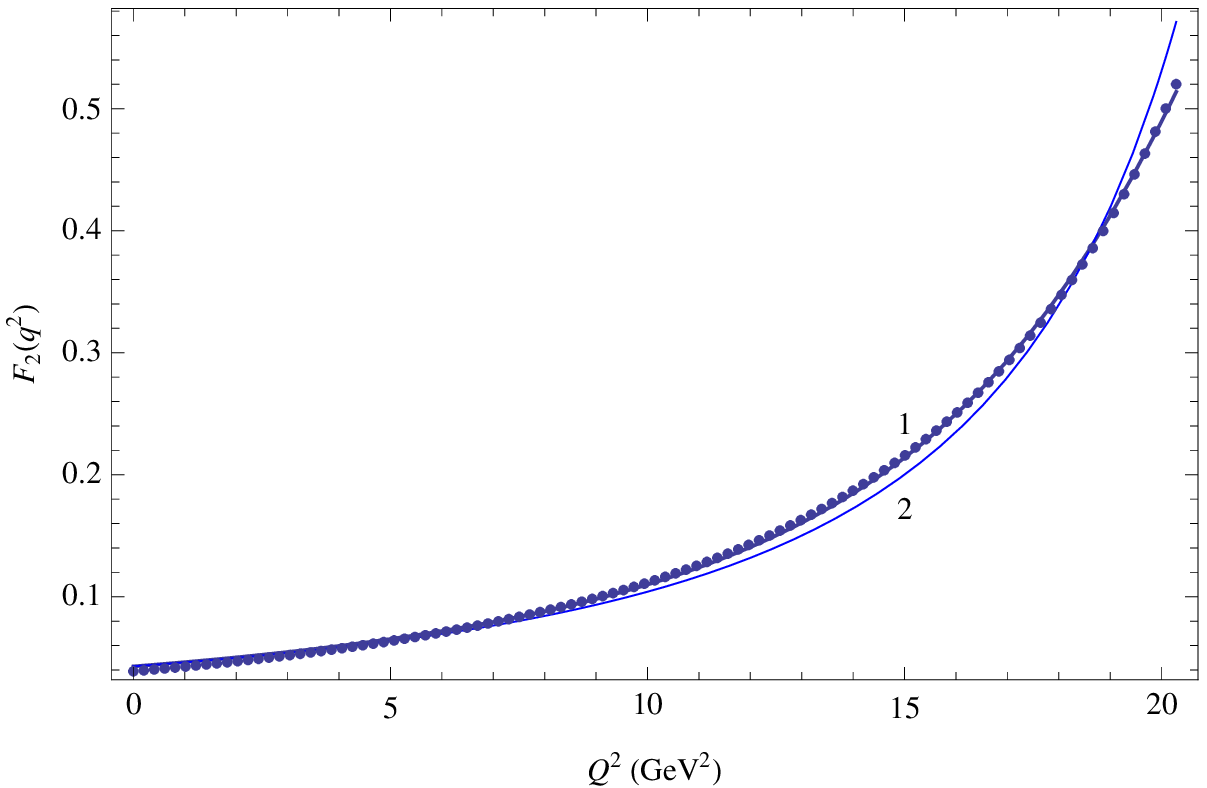,scale=.65}
\end{center}
\vspace*{-.6cm}
\noindent
\caption{Form factor $F_2^{V}$: exact result (dotted line), \\{}
double-pole approximation (curve 1) and 
dipole approximation (curve 2).
\label{fig5}}   

\begin{center}
\epsfig{figure=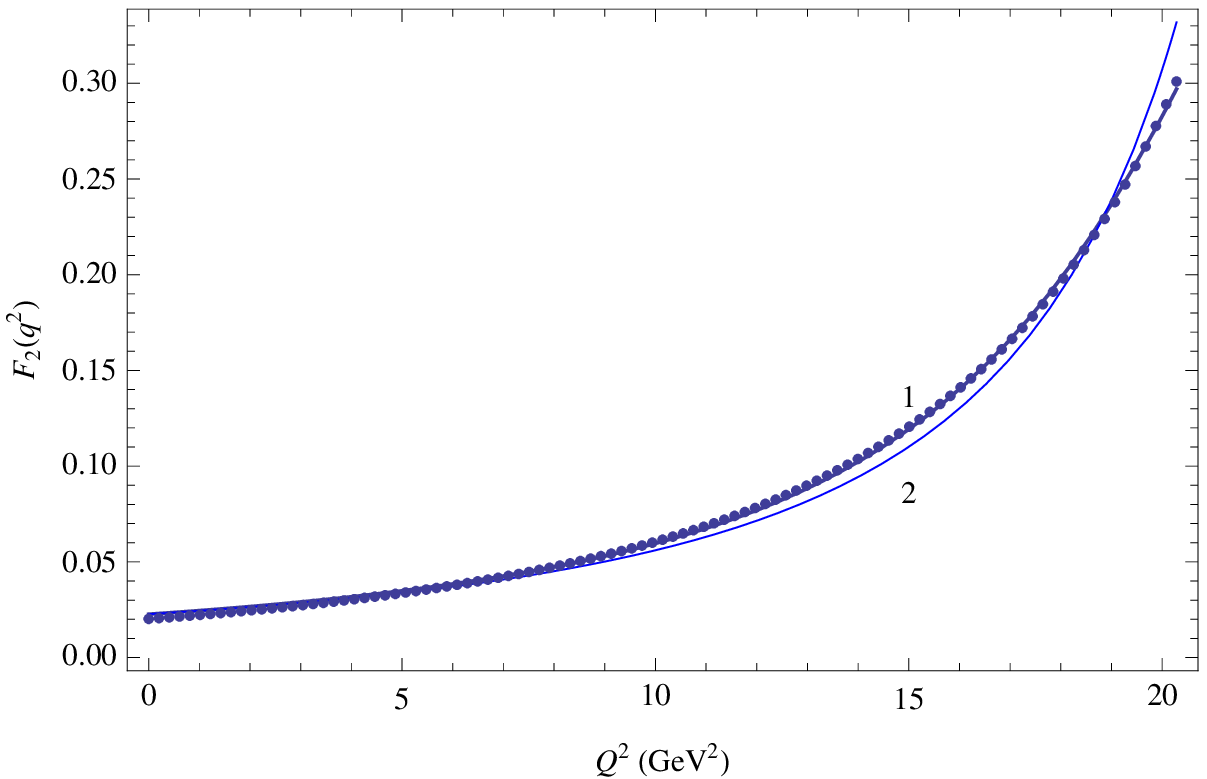,scale=.65}
\end{center}
\vspace*{-.6cm}
\noindent
\caption{Form factor $F_2^{V, {\rm HQL}}$: exact result (dotted line), \\{}
double-pole approximation (curve 1) and 
dipole approximation (curve 2).
\label{fig6}}   

\end{figure}

\clearpage 

\begin{figure}

\begin{center}
\epsfig{figure=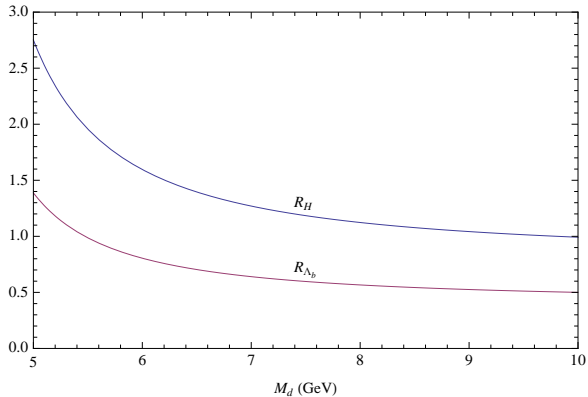,scale=.65}
\end{center}
\vspace*{-.6cm}
\noindent
\caption{Dependence of the ratios $R_{\cal H}$ and $R_{\Lambda_b}$  
on the dipole mass $M_d$ for the nonleptonic 
$\Lambda_b \to \Lambda + J/\psi,\psi(2S)$ transitions.   
\label{fig7}}   
\end{figure}

\begin{figure}[htb]
\begin{center}
\epsfig{figure=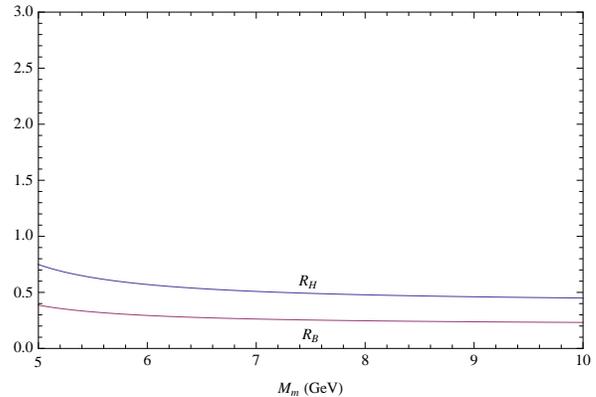,scale=.65}
\end{center}
\vspace*{-.6cm}
\noindent
\caption{Dependence of the ratios $R_{\cal H}$ and $R_{B}$ 
on the monopole mass $M_m$ for the nonleptonic $B \to K + J/\psi,\psi(2S)$ 
transitions.  
\label{fig8}}   
\end{figure}

\end{document}